\documentclass[aps,showpacs,twocolumn,floatfix]{revtex4}
\usepackage{epsfig}
\usepackage{dcolumn}
\usepackage{bm}

\begin{document}

\title{Quark models of dibaryon resonances in nucleon-nucleon scattering}

\author{J.L. Ping$^1$, H.X. Huang$^1$, H.R. Pang$^2$, Fan Wang$^3$
and C.W. Wong$^4$}

\affiliation{$^1$Department of Physics, Nanjing Normal University,
Nanjing, 210097, P.R. China \\
$^2$Department of Physics, Southeast University, Nanjing, 210094,
P.R. China \\
$^3$Department of Physics, Nanjing University, Nanjing, 210093, P.R.
China\\
$^4$Department of Physics and Astronomy, University of California, Los Angeles, CA 90095-1547, USA
}

\begin{abstract}
We look for $\Delta\Delta$ and $N\Delta$ resonances by calculating
$NN$ scattering phase shifts of two interacting baryon clusters of
quarks with explicit coupling to these dibaryon channels. Two phenomenological 
nonrelativistic chiral quark models giving similar low-energy $NN$
properties are found to give significantly different dibaryon resonance 
structures. In the chiral quark model (ChQM), the dibaryon system does 
not resonate in the $NN$ $S$-waves, in agreement with the experimental 
SP07 $NN$ partial-wave scattering amplitudes. In the quark delocalization
and color screening model (QDCSM), the $S$-wave NN resonances disappear
when the nucleon size $b$ falls below 0.53 fm. Both quark 
models give an $IJ^P = 03^+$ $\Delta\Delta$ resonance. At $b=0.52\,$fm, 
the value favored by baryon spectrum, the resonance mass is 
2390 (2420) MeV for the ChQM with quadratic 
(linear) confinement, and 2360 MeV for the QDCSM. Accessible from the 
$^3D_3^{NN}$ channel, this resonance is a promising candidate for the 
known isoscalar ABC structure seen more clearly in the 
$pn$$\rightarrow $$d\pi\pi$ production cross section at 2410 MeV in 
the recent preliminary data reported by the CELSIUS-WASA Collaboration. 
In the isovector dibaryon sector, our quark models give a bound or almost 
bound $^5S_2^{\Delta\Delta}$ state that can give rise to a $^1D_2^{NN}$
resonance. None of the quark models used has bound $N\Delta$ 
$P$-states that might generate odd-parity resonances.
\end{abstract}


\maketitle

\section{Introduction}
Quantum chromodynamics (QCD) is widely accepted as the fundamental
theory of the strong interaction.  Lattice QCD methods have recently been 
used to study low-energy hadronic interactions, including the 
nucleon-nucleon ($NN$) interaction \cite{Savage06}. However, QCD-inspired 
quark models are still the main tool for detailed studies of the 
baryon-baryon interaction.

The phenomenological quark model most commonly used in the study of
$NN$ interaction is the nonrelativistic chiral quark model (ChQM)
\cite{chiralmodel1,chiralmodel2,chiralmodel3}. Nonrelativistic 
kinematics makes the many-body treatment of the multiquark system
manageable, with the very convenient choice that the light quark mass
$m_q$ is just one third of the nucleon mass. With quarks, one needs a 
confinement potential to reproduce a distinctive QCD property, and a 
one gluon exchange (OGE) to account for
$\Delta$$-$$N$ mass difference and other details of baryon excitations.
The inclusion of pion exchange will take care of long-range baryon-baryon
interactions as well as some features in baryon structure, and is the 
consequence of the relative weakness of chiral symmetry breaking. 
Finally, scalar exchange is used to describe an extra 
intermediate-range attraction needed in nuclear forces. 
No other meson exchanges are included. ChQMs with a few chosen and a few 
adjusted parameters are able to give a surprisingly simple if only 
semi-quantitative picture of both baryon spectra and 
baryon-baryon interactions at relatively low energies. It is therefore of 
some interest to understand some of the limitations of 
these simple quarks models. 

The most problematic term in the ChQM is the scalar exchange term. It 
takes into account neglected channels containing $\Delta$s and pions, and 
should therefore vary when more of these channels are explicitly included 
in the coupled-channel calculation. Its effects always include the
exchange of two pions, and are called correlated if the two pions also 
interact with each other. Modern treatments of correlated two-pion exchange
show that in addition to a long-range scalar-isoscalar attraction
traditionally associated with scalar exchange, there is also a strong 
scalar-isoscalar repulsive core \cite{sigma}, a complication that has not 
yet been included in the ChQM. 

In the quark model, the forces between baryonic clusters of quarks and
antiquarks are like molecular forces between molecules of atoms built up from 
the forces between their constituents. This molecular model of nuclear forces 
has been extensively developed by using the quark delocalization and color 
screening model (QDCSM) \cite{QDCSM1}. In this model, quarks confined in one 
baryon are allowed to delocalize to a nearby baryon and to change the dynamics 
of the baryon-baryon interaction through a reduction of the confinement 
potential called color screening. The delocalization parameter that appears 
is determined by the dynamics of the interacting quark system, thus making it 
possible for the quark system to reach a more favorable configuration through 
its own dynamics in a larger Hilbert space. The model has been successfully
applied to $NN$ and hyperon-nucleon scatterings. The important
intermediate-range $NN$ attraction is achieved in this model by the
mutual distortion of the interacting nucleons, in a way that is very
similar to the mutual distortion of interacting molecules.

The main difference between the ChQM and the QDCSM is the mechanism
for intermediate-range attraction. Recently, we showed that both
the ChQM containing the $\sigma$-meson and the QDCSM
without it gave a good description of the low-energy $NN$ $S$- and
$D$-wave scattering phase shifts and the properties of deuteron with
almost the same quark-model parameters \cite{Chen_NN}. Thus the
$\sigma$-meson exchange can effectively be replaced by the quark
delocalization and color screening mechanism. It is not clear however
if their equivalence persists to higher energies where nucleons
overlap more strongly and baryon excitation and multiquark effects
become more important. 

Interest in multiquark system has persisted since R. Jaffe predicted 
the H-particle in 1977 \cite{Jaffe}.  All quark models, including those 
using lattice QCD techniques, predict that in addition to the
$q\bar{q}$ mesons and $q^{3}$ baryons, there should be multiquark
systems $(q\bar{q})^{2}$, $q^{4}\bar{q}$, $q^{6}$, quark gluon
hybrids $q\bar{q}g$, $q^{3}g$, and glueballs \cite{exotic}. Up to now
there has been no well established experimental candidate of these
multiquark states \cite{Close07}. Recently, the CELSIUS-WASA
Collaboration has reported preliminary results on the ABC anomaly in
the production cross section of the $pn$$\rightarrow
$$d\pi^{0}\pi^{0}$ reaction that suggests the presence of an
isoscalar $J^P=1^+$ or $3^+$ subthreshold $\Delta\Delta$ resonance,
with resonance mass estimated at $\sim 2410$ MeV and a width of $<
100$ MeV \cite{WASA07}. The relatively large binding energy involved
gives an object that is much closer to these interesting multiquark
states than a loosely bound system like the deuteron. Nonrelativistic 
quark models such as ChQMs and QDCSMs fitting both $N, \Delta$ 
masses and low-energy $NN$ scattering properties can be expected to 
give particularly interesting and parameter-free predictions for such
dibaryon multiquark states.

It thus appears worthwhile to extend our past calculation of $NN$
phase shifts to the resonance region near the $\Delta\Delta$ and
$N\Delta$ thresholds by including these excited dibaryon channels
in coupled-channel calculations. 
The $\Delta$ resonance is by far the most important low-energy baryon 
resonance. It dominates even the $\pi^-p$ cross sections where its production
is hindered relative to the production of isospin 1/2 $N^*$ resonances by a
factor of 2 from isospin coupling. In dibaryon channels, the 
$\Delta\Delta$ threshold at 2460 MeV is clearly separated from the 
$NN^*(1440)$ threshold at 2380 MeV, the second most interesting dibaryon 
threshold in this energy region. $\Delta\Delta$ bound states are of 
immediate interest in understanding resonance phenomena in this near
subthreshold energy region. The inclusion of $N^*(1440)$ would be of 
considerable interest in a broader study of dibaryon resonances, but its inclusion
is technically difficult for us because $N^*(1440)$ is commonly understood 
to be a monopole excitation of 
the nucleon that has a much more complicated quark wave function. In
contrast, our approximate description of the $N \rightarrow \Delta$ 
excitation as a simple spin-isospin excitation without any change in the 
radial wave function probably captures the essence of the physics involved. 
$\Delta$ excitations are thus within easy reach of the technology used in 
our previous coupled-channel calculations. Our first study in the 
resonance region will include only $\Delta$ excitations. This limitation 
of excited baryon degrees of freedom to $\Delta$s only has often been made 
in past studies of nuclear forces.~\cite{ExtendedBonn}

We shall use the same Salamanca ChQM \cite{Salamanca} and QDCSM used 
previously for $NN$ channels only \cite{Chen_NN} with additional sets 
of potential parameters to find out if their similarity persists into 
the resonance region. We are interested in particular in discovering 
how similar these simple quark models are in describing theoretical 
dibaryon resonances originating from $\Delta\Delta$ or $N\Delta$ bound
states when the $\Delta$s are treated as stable particles. In other
words, these dibaryon resonances are theoretical compound dibaryon 
states that are allowed 
to decay only via the $NN$ channels. A brief description of these two 
quark models of the baryon-baryon interaction is given in Section II. 

The $NN$ phase shifts for these quark-quark potentials are calculated 
using a coupled-channel resonating group formalism \cite{RGM} that includes
$\Delta\Delta$, $N\Delta$ and sometimes hidden-color channels as well.
No explicit pionic channels are included as the $\Delta$s are treated
as stable particles. In this context, a promising dibaryon resonance is 
taken to be one arising from a bound state below the $\Delta\Delta$ or 
$N\Delta$ threshold. The calculated results, including resonance masses 
and widths (FWHMs), are given in section III in those partial waves (PWs) 
where a theoretical dibaryon resonance appears in at least one quark model.
No $N\Delta$ bound state is found in any isovector odd-parity state in all
our quark models.

In Sect. IV, these results are compared to partial-wave analyses of 
$NN$ scattering amplitudes \cite{SP07,FA91}, where the presence of a
dibaryon resonance causes a rapid counterclockwise motion in
the Argand diagram. The possibility that the ABC effect in the
$pn$$\rightarrow $$d\pi\pi$ reaction is an isoscalar $NN$ resonance 
is also discussed. 

Section V contains brief concluding remarks on what we have learned
about quark dynamics in the $NN$ resonance region.

\section{Two quark models of baryon-baryon interactions}

\subsection{Chiral quark model}

The Salamanca ChQM is representative of chiral quark models. It has also been 
used to describe both hadron spectroscopy and nucleon-nucleon interactions. 
The model details can be found in \cite{Salamanca}. Only the Hamiltonian and 
parameters are given here.

The ChQM Hamiltonian in the baryon-baryon sector is
\begin{widetext}
\begin{eqnarray}
H &=& \sum_{i=1}^6 \left(m_i+\frac{p_i^2}{2m_i}\right) -T_c
+\sum_{i<j} \left[
V^{G}(r_{ij})+V^{\pi}(r_{ij})+V^{\sigma}(r_{ij})+V^{\rho}(r_{ij})+V^{C}(r_{ij})
\right],
\label{eq:H} \\
V^{G}(r_{ij})&=& \frac{1}{4}\alpha_s \boldsymbol{\lambda}_i \cdot
\boldsymbol{\lambda}_j
\left[\frac{1}{r_{ij}}-\frac{\pi}{m_q^2}\left(1+\frac{2}{3} 
\boldsymbol{\sigma}_i\cdot \boldsymbol{\sigma}_j \right)
\delta(r_{ij})-\frac{3}{4m_q^2r^3_{ij}}S_{ij}\right]+V^{G,LS}_{ij},
\nonumber \\
V^{G,LS}_{ij} & = & -\frac{\alpha_s}{4}\boldsymbol{\lambda}_i
\cdot \boldsymbol{\lambda}_j \frac{1}{8m_q^2}\frac{3}{r_{ij}^3}
[{\mathbf r}_{ij} \times ({\mathbf p}_i-{\mathbf p}_j)] 
\cdot(\boldsymbol{\sigma}_i + \boldsymbol{\sigma}_j),
\nonumber \\
V^{\pi}(r_{ij}) &=& \frac{1}{3}\alpha_{ch}
\frac{\Lambda^2}{\Lambda^2-m_{\pi}^2}m_\pi \left\{ \left[ Y(m_\pi
r_{ij})- \frac{\Lambda^3}{m_{\pi}^3}Y(\Lambda r_{ij}) \right]
\boldsymbol{\sigma}_i \cdot \boldsymbol{\sigma}_j
+ \left[ H(m_\pi r_{ij})-\frac{\Lambda^3}{m_\pi^3}
H(\Lambda r_{ij})\right] S_{ij} \right\} 
\boldsymbol{\tau}_i \cdot \boldsymbol{\tau}_j, 
\nonumber \\
V^{\sigma}(r_{ij})&=& -\alpha_{ch} \frac{4m_u^2}{m_\pi^2}
\frac{\Lambda^2}{\Lambda^2-m_{\sigma}^2}m_\sigma \left[ Y(m_\sigma
r_{ij})-\frac{\Lambda}{m_\sigma}Y(\Lambda r_{ij})
\right]+V^{\sigma,LS}_{ij}, 
~~~~
 \alpha_{ch}= \frac{g^2_{ch}}{4\pi}\frac{m^2_{\pi}}{4m^2_u},
 \nonumber \\
V^{\sigma,LS}_{ij} & = & -\frac{\alpha_{ch}}{2m_{\pi}^2}
\frac{\Lambda^2}{\Lambda^2-m_{\sigma}^2}m^3_{\sigma} 
\left[G(m_\sigma r_{ij})- \frac{\Lambda^3}{m_{\sigma}^3}G(\Lambda r_{ij})
\right] [{\mathbf r}_{ij} \times ({\mathbf p}_i-{\mathbf p}_j)]
\cdot(\boldsymbol{\sigma}_i + \boldsymbol{\sigma}_j),
\nonumber \\
V^{\rho}(r_{ij}) &=& \alpha_{chv} \frac{4m_u^2}{m_\pi^2}
\frac{\Lambda^2}{\Lambda^2-m_{\rho}^2}m_{\rho} 
\left\{ \left[Y(m_\rho r_{ij})-\frac{\Lambda}{m_\rho}Y(\Lambda r_{ij})
\right] + \frac{m_{\rho}^2}{6m_u^2}\left[ Y(m_\rho r_{ij})-
\frac{\Lambda^3}{m_{\rho}^3}Y(\Lambda r_{ij}) \right] 
\boldsymbol{\sigma}_i \cdot \boldsymbol{\sigma}_j \right. \nonumber \\
& & \left. -\frac{1}{2}\left[ H(m_\rho
r_{ij})-\frac{\Lambda^3}{m_\rho^3} H(\Lambda r_{ij})\right] S_{ij}
\right\} \boldsymbol{\tau}_i \cdot \boldsymbol{\tau}_j + V^{\rho,LS}_{ij},
~~~~~~
 \alpha_{chv}= \frac{g^2_{chv}}{4\pi}\frac{m^2_{\pi}}{4m^2_u},
 \nonumber \\
V^{\rho,LS}_{ij} & = & -\frac{3\alpha_{chv}}{m_{\pi}^2}
\frac{\Lambda^2}{\Lambda^2-m_{\rho}^2}m^3_{\rho} \left[ G(m_\rho
r_{ij})- \frac{\Lambda^3}{m_{\rho}^3}G(\Lambda r_{ij}) \right]
[{\mathbf r}_{ij} \times ({\mathbf p}_i-{\mathbf p}_j)]
\cdot(\boldsymbol{\sigma}_i + \boldsymbol{\sigma}_j)
~\boldsymbol{\tau}_i \cdot \boldsymbol{\tau}_j,
\nonumber \\
V^{C}(r_{ij})&=& -a_c \boldsymbol{\lambda}_i \cdot \boldsymbol{\lambda}_j
(r^2_{ij}+V_0) + V^{C,LS}_{ij}, \nonumber
\\
V^{C,LS}_{ij} & = & a_c \boldsymbol{\lambda}_i \cdot \boldsymbol{\lambda}_j
\frac{1}{8m_q^2}\frac{1}{r_{ij}}\frac{dV^c}{dr_{ij}}[{\mathbf
r}_{ij} \times ({\mathbf p}_i-{\mathbf p}_j)] \cdot
(\boldsymbol{\sigma}_i + \boldsymbol{\sigma}_j),
~~~~~~
 V^c = r_{ij}^2, 
\nonumber \\
S_{ij} & = &  \frac{(\boldsymbol{\sigma}_i \cdot {\mathbf r}_{ij})
(\boldsymbol{\sigma}_j \cdot {\mathbf r}_{ij})}{r_{ij}^2}
- \frac{1}{3}~\boldsymbol{\sigma}_i \cdot \boldsymbol{\sigma}_j.
\label{eq:ChQM}
\end{eqnarray}
\end{widetext}
Here $S_{ij}$ is quark tensor operator, $Y(x)$, $H(x)$ and $G(x)$ are
standard Yukawa functions \cite{DZYW03} , $T_c$ is the kinetic energy of the center
of mass, $\alpha_{ch} $ is the chiral coupling constant, determined 
as usual from the $\pi$-nucleon coupling constant. An additional $\rho$
meson exchange potential $V^\rho$ between quarks has been added to 
give an improved treatment of baryon-baryon $P$-states. Its parameters
will be specified in Sect.~\ref{sect:Results}. All other
symbols have their usual meanings.

Table~\ref{tab:ModelParameters} gives the model parameters used. For 
each set of parameters, the nucleon size $b$ that appears in 
Eq.(\ref{1q}) is given a pre-determined value. Two of the parameters 
($a_c, \alpha_s$) are fitted to $\Delta$$-$$N$ mass difference 
($1232-939\,$MeV) 
and the equilibrium condition for the nucleon mass at the chosen $b$. 
The absolute nucleon mass is controlled by a constant term $V_0$ in the 
confinement potential that does not affect the baryon-baryon interaction.
In ChQM2 \cite{ChQM2}, the deuteron binding energy (2.22 MeV) is fitted by 
varying the combination $m_\sigma, \Lambda$ calculated with the standard two 
$NN$ coupled channels $^3S_1$ and $^3D_1$ (called 2NNcc in the following). 
The remaining parameters $m_q, m_\pi$ and $\alpha_{ch}$ are fixed at chosen 
values. The numeral 2 in the 
name ChQM2 refers to its quadratic confinement potential. The model ChQM1 
\cite{chiralmodel1} uses a linear confinement potential instead. The model 
ChQM2a differs from ChQM2 in fitting a different equilibrium nucleon size $b$ 
but for simplicity $m_\sigma, \Lambda$ are allowed to remain at the ChQM2 
values. Its deuteron binding energy is reduced, but we do not consider the 
difference to be important in our study of dibaryon resonance properties.

\begin{table}
\caption{\label{tab:ModelParameters}  Parameters that differ in different 
models are given in this table. The dimension of each dimensional 
parameter is given within parentheses following each symbol: 
$b\,$(fm), $ a_c\,$(MeV\,fm$^{-2}$ if quadratic, but MeV\,fm$^{-1}$ if linear), 
$V_0\,$(fm$^2$) and $\mu\,$(fm$^{-2}$). Parameters having the same value
for all the quark models discussed in this paper are $m_q=313\,$MeV,
$m_\pi=138\,$MeV, $\Lambda/\hbar c=4.2\,$fm$^{-1}$, and 
$m_\sigma=675\,$(MeV) for ChMQs. The scattering 
length and effective range calculated for each potential 
are also given: $a_t, r_t$ for the triplet state $^3S_1$ and 
$a_s, r_s$ for the singlet state $^1S_0$, all in fm. The deuteron 
binding energy $\varepsilon_d\,$(MeV) is calculated from the
triplet effective-range parameters. }

\begin{tabular}{lccccc}\hline
             &  ChQM2  & ChQM2a & QDCSM0 & QDCSM1 & QDCSM3 \\
             & (ChQM1) &        &        &        &        \\
\hline\hline
$b$          & 0.518   &  0.60  &  0.48  & 0.518  &  0.60   \\
$a_c$        & 46.938  & 12.39  & 85.60  & 56.75  & 18.55   \\
             & (67.0)  &        &        &        &         \\
$V_0$        & -1.297  & 0.255  & -1.299 & -1.3598& -0.5279 \\     
$\mu$        &         &        &  0.30  &  0.45  &  1.00   \\
$\alpha_s$   & 0.485   & 0.9955 & 0.3016 & 0.485  &  0.9955 \\
$\alpha_{ch}$& 0.027   & 0.027  & 0.027  & 0.027  &  0.027  \\
             & (0.0269)&        &        &        &         \\
\hline
$a_t$        &  4.52   &  20.8  &  34.9  &  5.94  &  6.03  \\
$r_t$        &  1.56   &  2.24  &  2.27  &  1.75  &  1.67  \\
$\varepsilon_d$ &  3.35   &  0.11  &  0.04  &  1.75  &  1.64  \\
$a_s$        &  -170   &  -2.48 & -2.32  & -6.90  & -5.41  \\
$r_s$        &  2.17   &  5.42  &  4.48  &  2.63  &  3.56  \\
\hline
 
\end{tabular}

\end{table}

Finally, Table~\ref{tab:ModelParameters} also gives the effective-range (ER)
parameters of a 5-parameter ER formula in the $NN$ $^3S_1$ ($^1S_0$) state. 
The calculation uses with 5 (4) color-singlet channels denoted 5cc (4cc) and 
defined in the following section. The channels used include 2 $NN$ and 3 
$\Delta\Delta$ (an $NN$, 2 $\Delta\Delta$, and an $N\Delta$) channels. 
The deuteron binding energy 
$\varepsilon_d$ calculated from the two ER parameters of the 5cc calculation 
is also given in the table. According to \cite{BP06}, this approximation 
overestimates the binding energy, but by only 0.015 MeV. So the tabulated 
binding energies are sufficiently accurate for this qualitative study.
The same ER aproximation for a 2NNcc calculation gives 
$\varepsilon_d \approx 1.86\,$MeV for ChQM2, thus showing that the 3 
$\Delta\Delta$ channels increase $\varepsilon_d$ by about 1.5 MeV.

\subsection{Quark delocalization, color screening model}

The model and its extension were discussed in detail in
\cite{QDCSM1,QDCSM2}. Its Hamiltonian has the same form as Eq.(\ref{eq:H}),
but used with $V^{\sigma}=0$ and a different confinement potential
\begin{eqnarray}
V_{ij}^{CON}(r_{ij}) = -a_c \mbox{\boldmath{$\lambda_i \cdot \lambda_j$}}
 [f_{ij}(r_{ij}) + V_0], 
\end{eqnarray}
where $f_{ij}(r_{ij}) = r_{ij}^2$ if quarks $i,j$, each in the Gaussian 
single-quark wave function $\phi_\alpha$ of Eq.~(\ref{1q}), are on the 
same side of the dibaryon, i.e., both centered at ${\bf S}/2$ or at 
$-{\bf S}/2$. If quarks $i, j$ are on opposite sides of the dibaryon, 
\begin{equation}
f_{ij}(r_{ij}) = \frac{1}{\mu}\left(1 - e^{-\mu r_{ij}^2}\right).
\label{cs}
\end{equation}

Quark delocalization in QDCSM is realized by assuming that the single
particle orbital wave function of QDCSM as a linear combination of
left and right Gaussians, the single particle orbital wave functions
of the ordinary quark cluster model,
\begin{eqnarray}
\psi_{\alpha}({\bf S} ,\epsilon) & = & \left(
\phi_{\alpha}({\bf S})
+ \epsilon \phi_{\alpha}(-{\bf S})\right) /N(\epsilon), \nonumber \\
\psi_{\beta}(-{\bf S} ,\epsilon) & = &
\left(\phi_{\beta}(-{\bf S})
+ \epsilon \phi_{\beta}({\bf S})\right) /N(\epsilon), \nonumber \\
N(\epsilon) & = & \sqrt{1+\epsilon^2+2\epsilon e^{-S^2/4b^2}}; \nonumber \\
\phi_{\alpha}({\bf S}) & = & \left( \frac{1}{\pi b^2}
\right)^{3/4}
   e^{-\frac{1}{2b^2} ({\bf r}_{\alpha} - {\bf S}/2)^2} \nonumber \\
\phi_{\beta}(-{\bf S}) & = & \left( \frac{1}{\pi b^2}
\right)^{3/4}
   e^{-\frac{1}{2b^2} ({\bf r}_{\beta} + {\bf S}/2)^2}. 
\label{1q}
\end{eqnarray}
Quak delocalization with color screening is an approximate way of 
including hidden-color (h.c.) effects. 

The color screening constant $\mu$ in Eq.(\ref{cs}) is determined by fitting 
deuteron properties. The parameters of the QDCSM$i\,(i=1,3)$ used here are 
those of Set $i$ of \cite{Chen_NN}, and are given again in
Table~\ref{tab:ModelParameters}. For QDCSM0, $\mu$ is an estimated value
that has not been fine tuned to the deuteron binding energy. 
These models differ in the equilibrium nucleon size $b$.

\section{Results}
\label{sect:Results}

$NN$ scattering phase shifts are calculated  for the quark models of 
Table~\ref{tab:ModelParameters} to energies beyond the $\Delta\Delta$ or
$N\Delta$ threshold for different choice of coupled channels. We include
channels containing one or more $\Delta$s treated as stable particles, 
and channels containing hidden-color (h.c.) states. The resonating-group 
method (RGM), described in more details in \cite{RGM}, is used.

Past experience has suggested that reliable estimates of resonance 
masses can be made using non-decaying $\Delta$s \cite{chiralmodel1}. 
For the theoretical $N\Delta$ resonance $d'$ ($IJ^P = 20^-$) 
at the theoretical mass 2065 MeV (and an $N\Delta$ binding energy of 106 
MeV), an increase in the imaginary part of the $\Delta$ resonance energy 
by $10-15$ MeV is known to increase the $d'$ mass by only a few tenth of 
an MeV \cite{VGF96}. This result suggests that the complete neglect of 
the imaginary part $\Gamma/2 \approx 60\,$MeV of the $\Delta$ resonance 
energy will underestimate the $d'$ resonance mass by perhaps 2 MeV. 
If this result holds generally for other resonances at other binding  
energies, our estimates of the resonance masses can be expected to be 
good to a few MeV.

The use of coupled channels containing stable $\Delta$s does mean that the 
calculated $NN$ phase shifts do not describe inelasticities correctly. 
Thus they cannot be compared quantitatively to experimental phase 
parameters above the pion-production threshold in $NN$ channels with 
strong inelasticities. For this reason, the primary emphasis of this paper 
is the extraction of resonance energies from phase shifts 
in the resonance region. 

Our theoretical dibaryons are made up of two stable constituents below
their breakup threshold and are therefore real resonances in the model.
They have finite widths that come from the coupling to open $NN$ channels.

\begin{table}
\caption{\label{tab:3D3} The $\Delta\Delta$ or resonance mass $M$
and decay width $\Gamma$, in MeV, in five quark models for the
$IJ^P=03^+$ states. The channels included are one channel or 1c 
($^7S_3^{\Delta\Delta}$ only), two coupled channels or 2cc (1c $+\,^3D_3^{NN}$), 
4cc (2cc $+\,^{7,3}D_3^{\Delta\Delta}$), and 10 coupled channels as described in 
the text. The pure $^7S_3^{\Delta\Delta}$ bound state mass for ChQM1 is 
2456 MeV \cite{chiralmodel1}. }

\begin{tabular}{lcccccccccc}\hline
$N_{ch}$ & \multicolumn{2}{c}{ChQM2} &  \multicolumn{2}{c}{ChQM2a} &
\multicolumn{2}{c}{QDCSM0} & \multicolumn{2}{c}{QDCSM1}  & \multicolumn{2}{c}{QDCSM3} \\
           & $M$  & $\Gamma $ & $M$ & $\Gamma $ & $M$ & $\Gamma$ & $M$ 
                  & $\Gamma$  & $M$ & $\Gamma $\\ \hline\hline
1c         & 2425 &  -  & 2430 &  -  & 2413 & -  & 2365 &  - & 2276  & - \\
2cc        & 2428 & 17  & 2433 & 10  & 2416 & 20 & 2368 & 20 & 2278  & 19 \\
4cc        & 2413 & 14  & 2424 &  9  & 2400 & 14 & 2357 & 14 & 2273  & 17 \\
10cc       & 2393 & 14  &      &     & -    & -  & -    & -  & -     & -  \\
10cc$^{'}$ & 2353 & 17  &      &     & -    & -  & -    & -  & -     & -  \\
10cc$^{''}$& 2351 & 21  &      &     & -    & -  & -    & -  & -     & -  \\
\hline
\end{tabular}

\end{table}

\subsection{{\em I=0} states}

\begin{figure}
\includegraphics[width=3.4in]{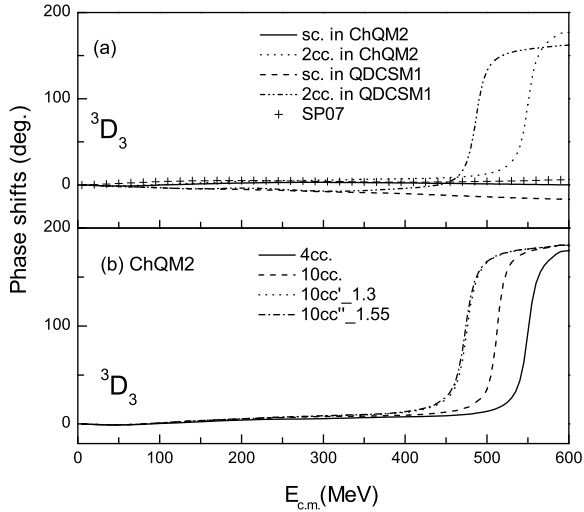}
\caption{\label{fig:3D3}
(a) $^3D_3^{NN}$ phase shifts calculated for a single channel (sc) and
two coupled channels (2cc) ($^3D_3^{NN}$ + $^7S_3^{\Delta\Delta}$) in
two different quark models (ChQM2 and QDCSM1) as functions of the c.m.
kinetic energy $E_{\rm cm} = W - 2M_N$, where $W$ is the total c.m.
energy and $M_N$ is the nucleon mass. The experimental phase shifts
of the partial-wave solution SP07 \cite{SP07} are also shown.
(b)  $NN$ $^3D_3$ phase shifts calculated in the ChQM2 using different
numbers of coupled channels, as described in more detail in the text. }
\end{figure}

Calculational details and results for the $IJ^P=03^+$ states are given 
in Table~\ref{tab:3D3}. The number of channels used in the theory is given
by $N_{ch}$. The theoretical pure $^7S_3^{\Delta\Delta}$ binding energy is
next estimated by diagonalizing the Hamiltonian matrix for this state in a 
generator-coordinate representation where the average baryon-baryon 
separation is taken to be less than 6 fm (in order to keep the matrix
dimension manageably small). In this way, the pure $^7S_3^{\Delta\Delta}$ 
is found to be bound by $35-190$ MeV, with the ChQM2 
mass 60 MeV lower than the ChQM1 mass obtained by \cite{chiralmodel1}. 
Coupling to the $^3D_3^{NN}$ channel causes this bound state to change
into an elastic resonance where the phase shift, shown in
Fig.~\ref{fig:3D3}, rises through $\pi/2$ at a resonance mass that has
been shifted up by 3 MeV. The same small mass increase caused by the
coupling to the $NN$ continuum is seen in all $J^P$ states studied here.
The result shows that the mass shift is always dominated by the $NN$
scattering states below the pure bound state mass rather than those
above it.

The table next shows that on coupling to the two $^{7,3}D_3^{\Delta\Delta}$
channels above the pure $^7S_3^{\Delta\Delta}$ bound state, the resonance
is pushed down in mass, as expected. The effect is not large, however,
being 15 MeV in ChQM2 and 11 MeV in QDCSM1, which has the same $\alpha_s$.

Both the pure $^7S_3^{\Delta\Delta}$ bound state and the associated
$^3D_3^{NN}$ resonance are lower in mass in QDCSM1, by about 60 MeV, 
than in ChQM2. Since the QDCSMs contain h.c. effects, we must
first determine how much h.c channels contribute to this mass 
difference. This study is done with ten channels (or case 10cc). 
Besides the four baryon channels shown in Table~\ref{tab:3D3}, they 
include the following six h.c. channels:
four $^3D_3$ channels of $^2\Delta_{8}\,^2\Delta_{8}$,
$^4N_{8}\,^4N_{8}$, $^4N_{8}\,^2N_{8}$, $^2N_{8}\,^2N_{8}$, the
$^7S_3(^4N_{8}\,^4N_{8})$ and the $^7D_3(^4N_{8}\,^4N_{8})$.  Here
the baryon symbol is used only to denote the isospin so that
$^2\Delta_{8}$ means the $T,S = 3/2, 1/2$ color-octet state. The
table shows that these six h.c. channels lower the ChQM2 resonance
mass by 20 MeV. Assuming that the QDCSMs account adequately for
h.c. contributions, we see that the QDCSM1 dibaryon resonance mass is now
lower than that in ChQM2 by only 36 MeV. 

It is interesting that the pure $^7S_3^{\Delta\Delta}$ bound state masses 
in the ChQM2 and ChQM2a of different nucleon size $b$ differ by only 5 MeV. 
In contrast, both bound state and resonance masses change by 90 MeV in the 
two QDCSMs with the same difference in the nucleon size $b$. This shows that 
the QDCS mechanism depends 
sensitively on the nucleon size. The possibility of a $^7S_3^{\Delta\Delta}$
bound state giving rise to a subthreshold resonance has been suggested 
previously \cite{dd003}, but the resonance masses for the present quark 
models are higher. 

\begin{figure}
\includegraphics[width=3.4in]{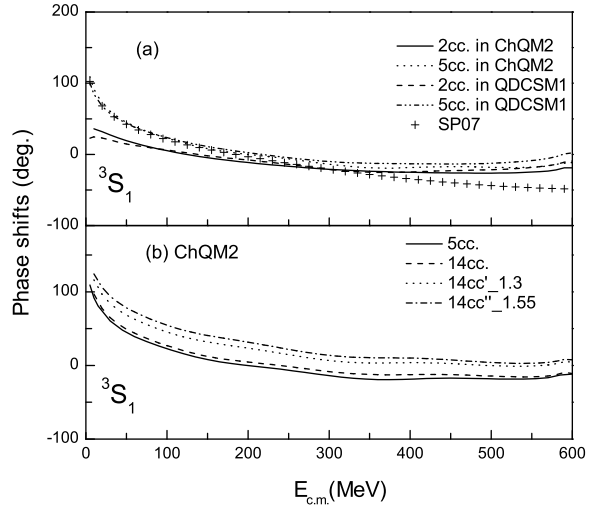}
\caption{\label{fig:3S1}
(a) $^3S_1^{NN}$ phase shifts calculated with two or five coupled
color-singlet channels defined in Table~\ref{tab:3S1} in two different
quark models ChQM2 and QDCSM1.
(b) $^3S_1^{NN}$ phase shifts calculated in the ChQM2 using different
numbers of coupled channels, as described in more detail in the text.  }
\end{figure}

The calculated $^3D_3^{NN}$ phase shifts are shown in Fig.~\ref{fig:3D3}. 
They differ noticeably between two quark models of very similar 
parameters that differ only in the replacement of the scalar exchange in 
ChQM2 by the QDCS mechanism in QDCSM1. 
The ChQM2 phase shifts agree better with experiment 
than the QDCSM1 phase shifts for $100 < E_{\rm cm} < 400\,$MeV. 
The situation is different in the $^3S_1^{NN}$ state shown 
in Fig.~\ref{fig:3S1} where the phase shifts from these two quark models 
agree, as already pointed out by \cite{Chen_NN}. 

Dibaryon resonance masses can be quite sensitive to short-range 
dynamics. For example, the ChQM2 resonance mass can be forced down to 
the lower QDCSM1 value by artificially increasing the attractive 
confinement interaction involving h.c. configurations that appear only 
for overlapping baryons. The change can be made in two different ways:
Either (1) increase the color confinement interaction strength among
h.c. channels, and also that between color-singlet channels and h.c.
channels (a model denoted 10cc') by an overall factor of 1.3;
or (2) increase the color confinement interaction between color-singlet
channels and h.c. channels only (model 10cc'') by an overall
factor 1.55. Fig.~\ref{fig:3D3}(b) shows that the resulting $^3D_3^{NN}$
phase shifts change noticeably only for $E_{\rm cm} > 350\,$MeV. 
In other words, only 
these ``high-energy'' $NN$ phase shifts are sensitive to changes in 
short-range dynamics when shielded by the strong centrifugal barrier in 
the $NN$ $D$-waves. Hence these changes in the confinement interactions, 
artificial as they are, are not excluded by the experimental phase shifts.

The resonance widths given in Table~\ref{tab:3D3} are FWHMs. They are
quite small, and agree with one another.

\begin{table}
\caption{\label{tab:3S1} The $\Delta\Delta$ or resonance mass $M$
and decay width $\Gamma$, in MeV, in five quark models for the
$IJ^P=01^+$ states. The channels included are 
1c ($^3S_1^{\Delta\Delta}$ only), 2cc (1c $+\,^3S_1^{NN}$), 
and 5cc (2cc $+\,^3D_1^{NN}+\,^{3,7}D_1^{\Delta\Delta}$). The pure 
$^3S_1^{\Delta\Delta}$ bound-state mass for ChQM1 is 2274 MeV 
\cite{chiralmodel1}. }

\begin{tabular}{lcccccc}\hline
$N_{ch}$ & ChQM2 & ChQM2a & QDCSM0 & QDCSM1 &
            \multicolumn{2}{c}{QDCSM3} \\
         & $M$  &  $M$ &  $M$ &  $M$ &  $M$ & $\Gamma$ \\ \hline\hline
1c       & 2366 & 2344 & 2317 & 2206 & 2115 & -  \\
2cc      & nr\footnotemark[1] & nr   & nr   & nr   & 2408 & 74 \\
5cc      & nr                 & nr   & nr   & nr   & 2393 & 70 \\ 
\hline
\end{tabular}
\footnotetext[1]{No resonance in these coupled channels.}
\end{table}

Results for the $IJ^P=01^+$ state are shown in Table~\ref{tab:3S1} and 
Fig.~\ref{fig:3S1}. The theoretical pure $^3S_1^{\Delta\Delta}$ state is 
bound by $100-350$ MeV, around twice the $^7S_3^{\Delta\Delta}$ binding energy. 
The coupling to the $^3S_1^{NN}$ channel has an unexpectedly large effect, 
pushing up the lowest of these bound $^3S_1^{\Delta\Delta}$ masses, by 293 MeV
in QDCSM3, so that it becomes a resonance at 2408 MeV. This very large mass 
shift is caused by the presence of a lower-mass state, the deuteron, in the 
admixed $^3S_1^{NN}$ channel. Admixing three additional channels with no
lower bound states pushes the resonance mass down a little to give for the 
5cc treatment the mass shift 
\begin{eqnarray}
\Delta M \equiv M_R(5cc) - M(1c) = 278\,{\rm MeV}.
\end{eqnarray}

The pure $^3S_1^{\Delta\Delta}$ bound state mass appears 100 MeV or more 
higher in the other four quark models. The additional large mass increase 
caused by the coupling to the $^3S_1^{NN}$ channel without or with the 
additional channels then pushes the state above the $\Delta\Delta$ threshold. 
Then no resonance appears. 

The pure $^3S_1^{\Delta\Delta}$ bound state mass in ChQM1 is higher by 160 
MeV than the QDCSM3 mass. So the $^3S_1^{NN}$ resonance also does not 
appear in ChQM1. We shall find that QDCSM3 model has an unusually rich 
dibaryon spectrum arising from an unusually strong attraction in 
$\Delta\Delta$ channels.

It is clear from Table~\ref{tab:3S1} that the strong $\Delta\Delta$
attraction in QDCSM3 is caused by the large nucleon size $b$ 
used there. As $b$ decreases, the $\Delta\Delta$ attraction also decreases.
This sensitivity to $b$ is not seen in ChQMs, thus showing that it is
caused by the QDCS mechanism.

In fact, the $^3S_1^{NN}$ resonance disappears somewhere between QDCSM1 
and QDCSM3. The critical value $b^{\rm crit}$ below which the 
resonance disappears can be estimated under the assumption that the mass 
shift $\Delta M$ caused by the coupling to the $NN$ channels is the same 
for all parameter sets. The critical point then appears when the bare
$^3S_1^{\Delta\Delta}$ bound-state mass is 2186 MeV. Interpolation from
the bound-state masses shown in Table~\ref{tab:3S1} gives
$b^{\rm crit} \approx 0.53\,$fm.

In contrast, the decrease of the pure $^3S_1^{\Delta\Delta}$ bound state 
mass for ChQM2s as $b$ increases from 0.52 fm to 0.60 fm is only 20\% that
of QDCSMs. Furthermore, $b$ can be increased to only 0.645 fm,
for the confinement potential strength $a_c$ turns negative above that value
and hence no ChQM2s can be constructed. At $b=0.645\,$fm, the pure 
$^3S_1^{\Delta\Delta}$ bound state mass is 2332 MeV, which is too large for 
the system to resonate on coupling to $NN$ channels. Hence ChQM2s have
no $^3S_1^{NN}$ resonance.

The pure $^3S_1^{\Delta\Delta}$ bound state mass for ChQM1 is 2274 MeV, 92 MeV 
below the ChQM2 mass. If its $b$ dependence is the same as ChQM2s, the decrease 
in bound state mass is insufficient, by about 50 MeV, to induce a resonance.

Turning now to the $NN$ phase shifts, Fig.~\ref{fig:3S1} shows that ChQM2 
and QDCSM1 give quite similar results, with the 5cc treatment giving 
much more attraction in both quark models, especially for 
$E_{\rm cm} < 150$ MeV.  There is fair agreement with experiment for 
$ E_{\rm cm} < 150\,$MeV, but all theoretical phase shifts become 
increasingly too attractive at higher energies. 

Fig.~\ref{fig:3S1}(b) shows the 14-channel $^3S_1^{NN}$ phase shifts
calculated in ChQM2 with the addition of nine h.c. channels:
eight $^3(S,D)_1$ channels of $^2\Delta_{8}\,^2\Delta_{8}$,
$^4N_{8}\,^4N_{8}$, $^4N_{8}\,^2N_{8}$, $^2N_{8}\,^2N_{8}$, and
$^7D_1{^4N_{8}\,^4N_{8}}$. Their inclusion causes the phase shift to
become only a little more attractive. 

The figure also shows the phase shifts obtained after the two 
arbitrary increases of the color confinement strength involving 
h.c. channels made previously for the $^3D_3^{NN}$ system. The phase 
shifts are now noticeably different from each other. Both are considerably
more attractive than those for case 14cc, but still not attractive enough 
at high energies for a resonance to appear below the $\Delta\Delta$
threshold. These large changes in the low-energy phase shifts are
inconsistent with experiment, thus showing that these additional h.c.
effects can now be excluded.

Two other $NN$ partial waves merit a short discussion. The 5cc $^3D_1^{NN}$ 
phase shifts, like their $^3S_1^{NN}$ partners, are nonresonant except for
QDCSM3. All theoretical phase shifts agree well with experiment, to around 
200 MeV. The quality of the theoretical $^3S_1^{NN}$ and $^3D_1^{NN}$ phase 
shifts shows that both quark models give good descriptions of the longer 
range part of the effective isoscalar central potential.
The isoscalar $^5S_2^{\Delta\Delta}$ is Pauli forbidden, while the pure 
$^3D_2^{\Delta\Delta}$ state is unbound. As a result, the 2cc $^3D_2^{NN}$ 
phase shifts are nonresonant. The QDCSM values agree quite well with 
experiment, while the ChQMs agree less well.

\subsection{{\em I=1} states}

\begin{table}
\caption{\label{tab:isovectors} The dibaryon or resonance mass and
decay width, all in MeV, in four quark models for $I=1$ states.
The channels included in the $J^P = 0^+$ state are 
1c ($^1S_0^{\Delta\Delta}$ only), 2cc (1c $+\,^1S_0^{NN}$), 
and 4cc (2cc $+\,^5D_0^{\Delta\Delta}+\,^5D_0^{N\Delta}$).
The channels included in the $J^P = 2^+$ state are 
1c ($^5S_2^{N\Delta}$ only), and 2cc (1c $+^1D_2^{NN}$). 
The bound-state masses for ChQM1 are 2304 MeV for $^1S_0^{\Delta\Delta}$
and 2171 MeV for $^5S_2^{N\Delta}$ \cite{chiralmodel1}.}

\begin{tabular}{lcccccccc}\hline
$N_{ch}$ & ChQM2 & ChQM2a & QDCSM0 & QDCSM1 & \multicolumn{2}{c}{QDCSM3} \\
 & $M$ & $M$ & $M$ & $M$ & $M$ & $\Gamma$ \\ \hline\hline
 & \boldmath{$J^P=0^+:$} & & & & & \\
1c  & 2395               & 2390 & 2335 & 2231 & 2148 & -  \\
2cc & nr\footnotemark[1] & nr   &  nr  &  nr  & 2448 & 106 \\
4cc & nr                 & nr   &  nr  &  nr  & 2433 & 128 \\ 
\hline
 & \boldmath{$J^P=2^+:$} & & & & & \\
1c & ub\footnotemark[2]  & ub   &  ub  &  ub  & 2167 & -  \\
2cc& nr                  & nr   &  nr  &  nr  & 2168 & 4  \\ \hline
\end{tabular}
\footnotetext[1]{No resonance in these coupled channels.}
\footnotetext[2]{Unbound.}
\end{table}

Table \ref{tab:isovectors} summarizes the results for two isovector
states with possible resonances. The $J^P = 0^+$ ($^1S_0^{NN}$)
state is qualitatively similar to the isoscalar $J^P = 1^+$
($^3S_1^{NN}$) state since they are mostly different spin states of the
same dibaryon pairs in the same relative orbital angular momenta. The pure 
$^1S_0^{\Delta\Delta}$ bound state in QDCSM3 is pushed up from its 
unperturbed energy of 2148 MeV by 300 MeV on coupling to the $^1S_0^{NN}$ 
channel by the presence of a lower-mass state, the well-known slightly unbound 
$^1S_0^{NN}$ state. The perturbed mass is still small enough for the system 
to resonate below the $\Delta\Delta$ threshold. 

In the remaining five quark models, the pure $\Delta\Delta$ mass is 
$80-250$ MeV higher. In each case, the strong coupling to the $NN$ channel 
pushes the state well into the $\Delta\Delta$ continuum, thus preventing a 
resonance from materializing. Following the procedure used for 
$^3S_1^{\Delta\Delta}$, the critical nucleon size below which the 
resonance disappears in the QDCSM is found to be $b^{\rm crit} \approx 0.56\,$fm.

\begin{figure}
\includegraphics[width=3.4in]{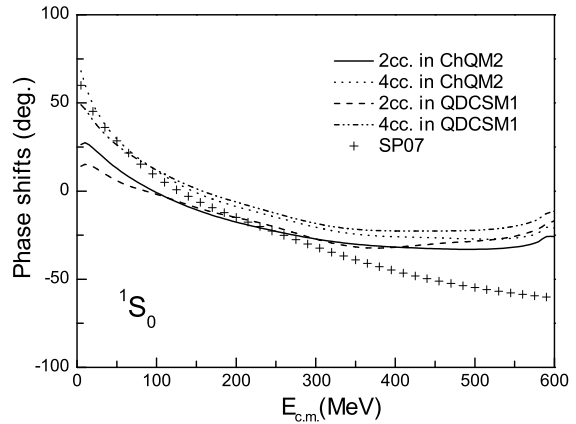}
\caption{\label{fig:1S0}
$NN$ $^1S_0$ phase shifts calculated with two or four coupled
color-singlet channels in two different quark models ChQM2 and QDCSM1. }
\end{figure}

Figure~\ref{fig:1S0} shows the non-resonant behavior of the coupled-channel 
$^1S_0^{NN}$ phase shifts for ChQM2 and QDCSM1. These phase shifts become
much more attractive than the experimental values from SP07 as the scattering 
energy increases into the resonance region, the effect being more than twice 
as strong as the similar behavior in the $^3S_1^{NN}$ state. All the quark 
models studied here do not give enough short-range repulsion in these 
$S$-states.

In the $J^P = 2^+$ state, the pure $^5S_2^{N\Delta}$ bound state appears at 
roughly the same mass straddling the $N\Delta$ threshold in all six quark 
models used. (Model differences in the $N\Delta$ $S$-state masses are much 
smaller for the high intrinsic spin state than for the low intrinsic spin 
states, just like the model differences in the $\Delta\Delta$ $S$-states.) The 
pure $^5S_2^{N\Delta}$ mass is pushed up only a little by coupling to the 
$^1D_2^{NN}$ continuum. It remains bound only in QDCSM3. The pure 
$^5S_2^{N\Delta}$ binding energy for ChQM1 is only 0.14 MeV. So the state 
is unlikely to remain below the $N\Delta$ threshold after coupling to the 
$NN$ channel. 

\begin{figure}
\includegraphics[width=3.2in]{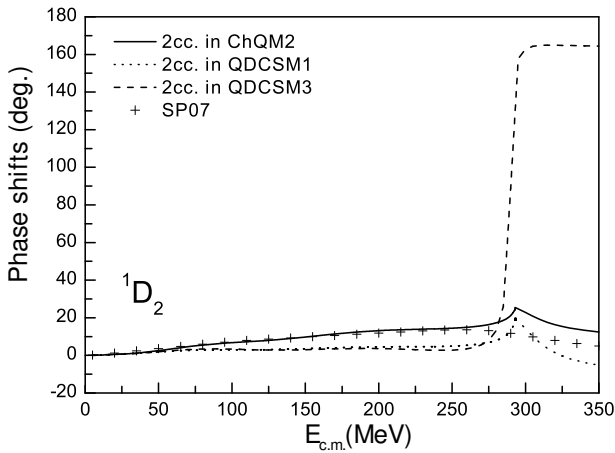}
\caption{\label{fig:1D2}
$NN$ $^1D_2$ phase shifts calculated with two coupled
color-singlet channels in three different quark models ChQM2, QDCSM1 
and QDCSM3. }
\end{figure}

The pure $N\Delta$ state is unbound in ChQM2. (It becomes bound if the 
attractive cental part of the scalar potential $V^\sigma$ shown in 
Eq.~(\ref{eq:ChQM}) is increased in strength by a multiplicative factor 
1.7.) The calculated non-resonant $NN$ phase
shifts are shown in Fig.~\ref{fig:1D2} for ChQM2 and QDCSM1. 
The prominent cusp is a threshold or Wigner cusp with its 
maximum located right at the $N\Delta$ threshold.
The phase shift above the threshold is the phase of $S_{11}$,
where the subscript 1 denotes the $NN$ channel. For comparison, the 
resonant phase shifts for QDCSM3 are also given.

Even though a resonance appears only in one quark model in our limited
theoretical treatment, the masses involved are sufficiently close to one 
another and to the $N\Delta$ threshold so that they describe similar 
dynamical situations to within the uncertainties of the models. Moreover, 
the large $\Delta$ width when included would cause the state to straddle 
the $N\Delta$ threshold for all these quark models. We therefore consider 
a $^5S_2^{N\Delta}$ resonance near the $N\Delta$ threshold to be possible
in all these quark models. 

In fact, inelastic Argand looping (which we shall define in
Sect.~\ref{sect:Discussion}) has been obtained by Entem, Fernandez 
and Valcarce \cite{EFV03} in the $^1D_2$ and possibly also $^3F_3$ systems 
for a ChQM having $\alpha_s = 0.4977$, only a little larger than the value 
0.485 used in ChQM1 or ChQM2. The crucial feature in their treatment is 
the explicit inclusion of $NN$ inelasticities by giving decay widths to 
the $\Delta$s appearing in the coupled-channel treatment. There exist 
quite extensive PW solutions of both $NN$ \cite{SP07} and $\pi d$ 
\cite{C500} scattering amplitudes in these and other isovector dibaryon 
systems. They could yield interesting information concerning quark dynamics 
in this resonance region. 

We do not find any resonance attributable to an $N\Delta$ or $\Delta\Delta$ 
bound state in any of the quark models in the following
four isovector states: (a) the $^3P_{0,1}, ^3F_3$ states (with each state
calculated using three coupled color-singlet channels of the same
quantum numbers for $NN$, $N\Delta$ and $\Delta\Delta$ constituents, 
respectively), and (b) the $J^P = 2^-$ state (using the four 
color-singlet channels $^3P_2$ of $NN$, $N\Delta$, $\Delta\Delta$, 
and $^7P_2^{\Delta\Delta}$).

\begin{figure}
\includegraphics[width=3.4in]{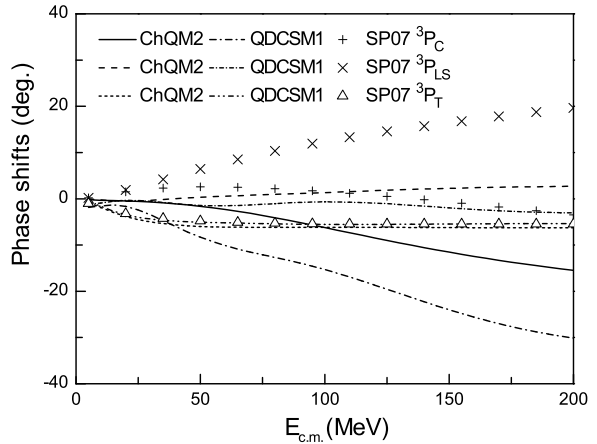}
\caption{\label{fig:Pwaves}
The central, spin-orbit and tensor components of the $^3P_J^{NN}$ phase shifts 
calculated with three coupled color-singlet channels (containing $NN$, $N\Delta$ and 
$\Delta\Delta$) in two different quark models ChQM2 and QDCSM1. }
\end{figure}

It is worthwhile to show in Fig.~\ref{fig:Pwaves} the difference between ChQM2 
and QDCSM1 in the well-known decomposition into central, spin-orbit and tensor 
components of the $^3P_J^{NN}$ phase shifts:
\begin{eqnarray}
^3P_C & = & \textstyle{\frac{1}{9}}\,^3P_0 + \textstyle{\frac{1}{3}}\,^3P_1 
+ \textstyle{\frac{5}{9}}\,^3P_2, \nonumber \\
^3P_{LS} & = & - \textstyle{\frac{1}{6}}\,^3P_0 - \textstyle{\frac{1}{4}}\,^3P_1 
+ \textstyle{\frac{5}{12}}\,^3P_2, \nonumber \\
^3P_T & = & - \textstyle{\frac{5}{36}}\,^3P_0 + \textstyle{\frac{5}{24}}\,^3P_1 
- \textstyle{\frac{5}{72}}\,^3P_2.
\end{eqnarray}
We see that the inclusion of one pion exchange (OPE) takes good care of the 
tensor component, but both quark models give too weak spin-orbit components
and too repulsive central components, especially for QDCSM1 
\cite{chiralmodel1,QDCSM1}.

As is well known \cite{chiralmodel1}, the problem with the spin-orbit 
component comes about because the included OPE potential though clearly needed
to generate a $NN$ tensor force also contributes to the $\Delta$$-$$N$ mass 
difference.  The color coupling constant $\alpha_s$ needed to account for this 
mass difference is then reduced to 0.3-1.0 from the value 1.7 in quark models 
without pion exchange. This weaker $\alpha_s$ gives in turn a weaker 
baryon-baryon spin-orbit potential from OGE. The additional spin-orbit 
contribution from the scalar exchange used in the ChQMs is not enough to 
compensate for the deficit. Resolution of this problem would require
significant modifications to the quark models used.

\begin{figure}
\includegraphics[width=3.4in]{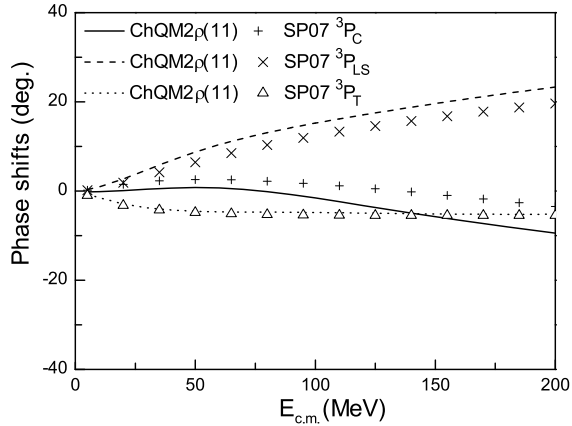}
\caption{\label{fig:ChQM2rho11Pwaves}
The central, spin-orbit and tensor components of the $^3P_J^{NN}$ phase 
shifts calculated with only a single uncoupled $NN$ channel in the quark model 
ChQM2$\rho$(11). }
\end{figure}

A simple way to study the spin-orbit problem in the present limited objective 
of looking for dibaryon resonances without overhauling these quark 
models is to just modify a term or add terms to the quark-quark interaction 
phenomenologically. A number of related quark models are thus generated: 
a modified ChQM2m model, where the one gluon exchange strength $\alpha_s$ of 
the spin-orbit potential $V_{ij}^{G,LS}$ has been increased 5 times, and a 
number of ChQM2$\rho$(f) models with the additional $\rho$ meson exchange 
potential $V^\rho$ displayed in Eq.~(\ref{eq:H}). Here 
\begin{eqnarray}
f = \frac{\alpha_{chv}}{\alpha_{chv0}} 
\end{eqnarray}
is the multiplicative increase of the $\rho$-quark-quark coupling strength 
above the usual and customary value $\alpha_{chv0} = 0.021$, corresponding 
to the coupling constant $g_{chv0} = 2.351$ used in \cite{DZYW03}. Since
the effects on these $P$-wave $NN$ phase shifts of the coupling to $N\Delta$ 
and $\Delta\Delta$ channels are quite small, we use only a single 
color-singlet $NN$ channel to calculate the $NN$ phase shifts for these 
modified models. Fig.~\ref{fig:ChQM2rho11Pwaves} shows that the 
resulting central, spin-orbit and tensor components of the $^3P_J^{NN}$ 
phase shifts for the quark model ChQM2$\rho$(11) give quite good agreement 
with the experimental SP07 values. By diagonalizing the Hamiltonian matrix
in the appropriate $N\Delta$ channel, we find that none of the modified quark 
models described in this paragraph has a pure $N\Delta$ $P$-wave bound state. 

The difficulty of forming $P$-wave bound states can readily be appreciated 
in the attractive square-well potential model. It is well known that to 
bind a $P$-state, its attractive potential depth must be four times that 
needed to bind an $S$-state \cite{Schiff68}. To see how far we are from 
$N\Delta$ $P$-wave bound states for our quark models, we increase the 
strength of the attractive central scalar potential $V^\sigma$ in ChQM2 
by a multiplicative factor $f_s$. The $^3P_2^{N\Delta}$ state then 
becomes bound at $f_s \gtrsim 3.1$, with a binding energy of 6 (22) MeV when 
$f_s = 3.2\: (3.4)$. The potentials in the other $^3P_J^{N\Delta}$ states 
are less attractive.

A similar situation holds for ChQM2$\rho$(11): Its $^3P_2^{N\Delta}$ 
($^5P_2^{N\Delta}$) state is unbound. It becomes bound only when its 
attractive central scalar potential strength has been increased by a 
multiplicative factor $f_s \approx 3.0$ (2.6). It is thus clear that 
$N\Delta$ $P$-wave bound states would appear only for quark models with 
substantially stronger attraction than those studied here.

Concerning the missing $P$-wave attraction in both ChQM2 and QDCSM1, one 
cannot just use the isoscalar scalar $\delta$ meson of mass 980 MeV that 
appears in the Bonn potentials \cite{Machleidt89} because of the cutoff 
mass $\Lambda \approx 830$\,MeV used in our chiral quark models. It is
not a trivial problem to reconcile these two classes of models for nuclear 
forces.

Turning now to our experimental knowledge of possible resonances in 
$NN$ $P$-states, we recall that in the PW analysis FA91 \cite{FA91}, 
resonance poles are found for the isovector odd-parity $NN$ states 
$^3P_2, ^3F_2$ and $^3F_3$. These empirical resonance-like solutions 
reproduce the empirical Argand loopings 
of the PW solutions, but many studies in the past \cite{LSS86} have left 
unresolved the question of whether these Argand loopings represent new 
dibaryon resonances. The difficulty centers around the observation of 
Brayshaw \cite{Brayshaw76} that when decay channels with three or more final 
particles are present, Argand looping can appear in models known to have no 
resonance because their S-matrix has no pole. Brayshaw has given an explicit 
dynamical example for the case of the $pp$ $^1D_2$ state coupled to $N\Delta$ 
and $\pi^+d$ channels. The strong energy dependence that causes the Argand 
looping in his model comes from a logarithmic rescattering singularity in 
the $N\Delta \leftrightarrow \pi^+d$ transition amplitude near the $N\Delta$ 
threshold. The physical situation this singularity describes is the 
oscillation or exchange of a nucleon between the decaying $\Delta$ and the 
second or spectator nucleon with which it forms the bound deuteron $d$. In
Brayshaw's model, there is no new $^1D_2^{pp}$ resonance near the 
$N\Delta$ threshold.

To complete our very brief review of resonance conditions, we should 
mention that recent studies of $\pi N$ resonances \cite{SSL08,WA08} 
have shown that the speed test can track the positions of resonance poles, 
if present, more reliably than the time delay criterion. (The speed test 
determines the resonance energy from the maximum speed of Argand looping
as a function of the on-shell kinetic energy, while the time delay test 
locates it by maximizing the positive time delay of the scattered wave 
packet relative to the free wave packet.) A different kind of complication
can appear in dynamical models that already have resonance poles, namely 
that the speed test can fail because the Argand looping does not have a 
solution with maximum speed \cite{SSL08}. This is an extension to another 
dynamical model (one having two coupled nonrelativistic two-body channels) 
of another old observation of Brayshaw that when relativistic many-body 
channels are present, true resonance poles can appear without any Argand 
looping \cite{Brayshaw76a}.

In view of all these complications, we shall take the tentative but 
conservative position that a promising dibaryon resonance is one involving 
at least one $\Delta$ or $N^*$ baryon that is a bound state below the 
dibaryon breakup threshold in the absence of a centrifugal potential when 
these baryons are treated as stable particles. In practice, the only excited
baryon we are able to describe with some degree of confidence is the 
$\Delta$. In the limited context of our quark models, we consider the 
$^1D_2^{NN}$ structure a promising dibaryon resonance, at least for some 
of our quark models, but not the $NN$ $P$-wave Argand loopings.

\section{Discussion}
\label{sect:Discussion}

\begin{table}
\caption{\label{tab:ResonanceSummary} Mass, decay widths (both in MeV)
and branching ratio of theoretical baryon resonances in five quark
models and comparison with partial-wave analyses of experimental data 
from SP07 \cite{SP07} and FA91\cite{FA91}. $M_R$ for ChQM2a is estimated 
from the 4cc value and is shown within parentheses. }

\begin{tabular}{lrrrr}\hline
$NN$           &  $^1S_0$  & $^3S_1$ & $^1D_2$ & $^3D_3$  \\
               &           & $^3D_1$ &         &          \\
$\Delta\Delta$ &  \boldmath{$^1S_0$}  & \boldmath{$^3S_1$}   &
& \boldmath{$^7S_3$} \\
               &  $^5D_0$  & $^{3,7}D_1$ &         & $^{3,7}D_3$ \\
$N\Delta$      &  $^5D_0$  &         & \boldmath{$^5S_2$} & \\
 \hline\hline
  & ChQM2: & & &   \\
$M_R$              &       &         &         &   2393   \\
$\Gamma_{NN}$      &       &         &         &     14   \\
$\Gamma_{\rm inel}$ &      &         &         &    136   \\
$B_{NN}$           &       &         &         &   0.09   \\
\hline
  & ChQM2a: & & &   \\
$M_R$              &       &         &         &  (2404)  \\
$\Gamma_{NN}$      &       &         &         &      9   \\
$\Gamma_{\rm inel}$ &      &         &         &    149   \\
$B_{NN}$           &       &         &         &   0.06   \\
\hline
  & QDCSM0: & & &   \\
$M_R$              &       &         &         &   2400   \\
$\Gamma_{NN}$      &       &         &         &     14   \\
$\Gamma_{\rm inel}$ &      &         &         &    144   \\
$B_{NN}$           &       &         &         &   0.09   \\
\hline
  & QDCSM1: & & &   \\
$M_R$              &       &         &         &   2357   \\
$\Gamma_{NN}$      &       &         &         &     14   \\
$\Gamma_{\rm inel}$ &      &         &         &     96   \\
$B_{NN}$           &       &         &         &   0.13   \\
\hline
$\alpha_s^{\rm crit}$ & 0.60 & 0.57  &         &          \\ 
\hline 
  & QDCSM3: & & &   \\
$M_R$              &  2433 &   2393  &   2168  &   2273   \\
$\Gamma_{NN}$      &   130 &     70  &      4  &     17   \\
$\Gamma_{\rm inel}$ &  190 &    136  &    117  &     33   \\
$B_{NN}$           &  0.41 &   0.34  &   0.03  &   0.34   \\
\hline
 & SP07: & & &  \\
$M_R$              & none  &  none? &  2148\footnotemark[1] &   ?   \\
$\Gamma $          &       &        &   118\footnotemark[1] &        \\
$B_{NN}$      &            &        &  0.29    &  0.26\footnotemark[2]\\
\hline

\end{tabular}
\footnotetext[1]{Pole position of FA91.}
\footnotetext[2]{At $W = 2400\,$MeV.}
\end{table}

The theoretical resonance properties calculated in the
last section are summarized in Table~\ref{tab:ResonanceSummary}. The
first line for each dibaryon type gives the dominant PW. The PW
responsible for the resonance trapping in the theory is shown in
bold type. The experimental information used for the comparison is
the PW solution SP07 \cite{SP07} of $NN$ scattering data. The four
states are arranged in order of increasing relative orbital angular 
momentum $\ell_{NN}$.

Each of our calculated resonances is an elastic resonance where the
scattering phase shift rises sharply through $\pi/2$. The Argand plot
of its complex PW amplitude
\begin{eqnarray}
T = \frac{S-1}{2i}
\label{T}
\end{eqnarray}
shows rapid counterclockwise motion on the unitarity circle. This
mathematical behavior describes a physical picture where the $NN$
system resonates or continues to ``sound'' due to its partial trapping
into the resonance region, namely the closed channel containing one or
two $\Delta$s. Each of these elastic resonances has a finite but very
small elastic (or $NN$) width. 

The resonance properties shown in the table must be corrected for the 
width of a decaying $\Delta$, leading to the appearance of pionic channels. 
Inelasticities cause the reduction $|S| < 1$ and the restriction of the 
Argand plot to the interior of the unitarity circle. We shall avoid 
Brayshaw's two complications from many-body channels by considering only 
channels where for stable $\Delta$s, the dibaryon system in our quark 
model treatment is a bound or almost bound two-body state. This restriction 
allows us to take the standard position that for these special states, 
rapid counterclockwise Argand looping is an acceptable signal of an
inelastic resonance~\cite{LSS86,KT83}. In physical terms, leakage
of the trapped system into pionic channels reduces the effect ``heard''
in the $NN$ channel but does not eliminate it altogether. From the 
perspective of the quark models used here, the $I=1$ odd-parity $NN$ 
Argand structures are not promising candidates for dibaryon resonances. 

We shall estimate inelasticities only at the crudest level of
branching ratios, with
\begin{eqnarray}
B_{NN} = \frac{\Gamma_{NN}}{\Gamma},
\label{BRNN}
\end{eqnarray}
where $\Gamma = \Gamma_{NN} + \Gamma_{\rm inel}$ depends on the
inelastic width $\Gamma_{\rm inel}$ caused by decaying $\Delta$s.
Close to the breakup threshold where the $\Delta$s are almost on-shell,
the inelastic width can be related approximately to the $\Delta$
width $\Gamma_{f\Delta} = 120\,$MeV in free space by only accounting
for the reduction in phase space available to a decaying bound
$\Delta$ whose mass has been reduced to roughly
$M_{b\Delta} \approx M_{f\Delta} - 0.5 B$, where $B$ is the binding
energy of the dibaryon. Then \cite{DLMS07}
\begin{eqnarray}
\Gamma_{b\Delta}(M_{b\Delta}) &\approx& \Gamma_{f\Delta}
\frac{k_b^{2\ell}\rho(M_{b\Delta})}{k_f^{2\ell}\rho(M_{f\Delta})},
\label{GammaBDel}
\end{eqnarray}
where $k$ is the pion momentum in the rest frame of the decaying 
$\Delta$, $\ell = 1$ is the pion angular momentum, and
\begin{eqnarray}
\rho(M) = \pi \frac{k E_\pi E_N}{M}
\label{rho}
\end{eqnarray}
is the two-body decay phase space at mass $M$ when each decay product
has c.m. energy $E_i, \,i = \pi,\,$N. We shall use this crude
estimate indiscriminately even far below the breakup threshold,
but the harm done is not great because most promising resonances
are near the threshold.

If each decaying $\Delta$ in the dibaryon has a Breit-Wigner (BW) 
distribution of width $\Gamma_{b\Delta}$, the total mass of two decaying 
$\Delta$s can be shown to have a BW distribution with width 
$2\Gamma_{b\Delta}$. Hence
\begin{eqnarray}
\Gamma_{\rm inel} &\approx&n_{b\Delta}\Gamma_{b\Delta}(M_{b\Delta}),
\label{GammaInel}
\end{eqnarray}
$n_{b\Delta}$ is the number of $\Delta$s in the resonance. The
results for different theoretical dibaryons are summarized in Table
\ref{tab:ResonanceSummary}. The shift in resonance masses caused by
the coupling to pionic channels has not been included in these
estimates.

The experimental branching ratios shown in Table~\ref{tab:ResonanceSummary}
are obtained from the energy-dependent SP07 PW solution \cite{SP07} at
the stated resonance masses using the formula for coupled PWs
\begin{eqnarray}
B_{NN} = \frac{\sigma_{{\rm el},J}}{\sigma_{{\rm tot},J}}
= \frac{|T_{11}|^2 + |T_{12}|^2}{{\it Im}T_{11}},
\label{BR-PWA}
\end{eqnarray}
while $T_{12} = 0$ for an uncoupled PW. Here the subscripts $i,j$ in 
$T_{ij}$ are channel labels.

Our theoretical estimates 
can now be compared to the experimental results from the PW analysis 
of $NN$ scattering data, in Table~\ref{tab:ResonanceSummary}. 
The absence in the SP07 PW solution of a resonance accessible from the 
$^1S_0^{NN}$ channel and its probable absence in the $^3S_1^{NN}$ 
give an approximate upper bound on the nucleon size of 
$b^{\rm crit} \approx 0.53\,$fm for the QDCSM. This upper bound
causes the $^3D_3^{NN}$ resonance mass to exceed 2340 MeV, 
the interpolated value for the QDCSM at $b = 0.53\,$fm. 

The $^3D_3^{NN}$ resonance appears also in both ChQM2 and ChQM2a at around 
2400 MeV, the resonance mass being not very sensitive to $b$. The resonance 
mass has not been calculated for the ChQM1, but is about 2420 MeV. 
$NN$ $S$-wave resonances do not appear in ChQM with quadratic or 
linear confinement.

Table~\ref{tab:ResonanceSummary} also shows that the theoretical 
$^3D_3^{NN}$ branching ratio $B_{NN}$ is smaller than the SP07 value. 
This means that the theoretical coupling to the trapping $\Delta\Delta$ 
channel is too weak. The theoretical branching ratio $B_{NN}$ for the QDCSM3 
resonance in the $^1D_2^{NN}$ state is much smaller than the SP07 value
except for QDCSM3. This suggests that the theoretical coupling to the 
trapping $N\Delta$ channel is also too weak.

\subsection{The ABC effect}

The ABC effect, named after Abashian, Booth and Crowe \cite{ABC60}
who first observed it, describes an enhancement above phase space of
the missing-mass spectrum of the inelastic fusion reaction
$pd$$\rightarrow $$^3{\rm He}X$. Subsequent experimental studies have been
reviewed recently by Clement {\it et al.} (or WASA07) \cite{WASA07},
who also report preliminary results for the exclusive reaction
$pd$$\rightarrow $$^3{\rm He}\pi\pi$ from the WASA Collaboration, and by
Bashkanov \cite{Bash06}.

The enhancement is associated with a $\pi\pi$
invariant mass $< 340\,$MeV, with the two pions emitted in parallel
and in relative $S$-wave opposite in direction to the recoiling
nucleus. The structure is isoscalar because it is seen in
$dd$$\rightarrow $$^4{\rm He}X^0$,
but not in $dp$$\rightarrow $$^3{\rm H}X^+$.

In the reaction $pd$$\rightarrow $$^3{\rm He}\pi^0\pi^0$,
the differential cross section
$d\sigma/dM_{\pi^0\pi^0}$ at fixed $M_3$, the invariant $^3{\rm He}\pi^0$
mass, has a maximum at $\approx 3080\,$MeV, just under the value of
$2M_N + M_\Delta \approx 3100\,$MeV. The cross section has a FWHM of 
about $\Gamma_3 \approx 130\,$MeV, the same as $\Gamma_{f\Delta}$. 
(These numbers are from Fig.~5-5 of \cite{Bash06}.) Hence the pion in
$M_3$ appears to have come from the decay of a slightly bound $\Delta$.
These features of the experimental data suggest that the ABC effect in
these reactions comes from the decay of a $\Delta\Delta$ bound state.
The preliminary WASA07 resonance peaks at 2410 MeV, about 50 MeV below
the $\Delta\Delta$ threshold, with a width $< 100\,$MeV. If the 
estimated width holds up, it would eliminate the larger values of
$2\Gamma_{b\Delta} \approx 160\,$MeV (from Table~\ref{tab:ABC}) to
$2\Gamma_3 \approx 260\,$MeV (from Fig.~5-5 of \cite{Bash06}). 
As pointed out by WASA07, such an outcome would disagree with the 
situation in the $^1D_2$ resonance whose width is close to the 
free $\Delta$ width even though the resonance straddles the $N\Delta$
breakup threshold.

\begin{table}
\caption{\label{tab:ABC} Experimental peak position (in MeV) and
production cross section
$\sigma_{d\pi\pi}$ (in mb) and width (in MeV) of the ABC effect
compared to the estimate from the partial-wave solution SP07 of
$np$ scattering amplitudes under the assumption that the ABC effect
comes from a dibaryon resonance in the specified PW at the resonance
mass 2410 MeV. These estimates are described in more detail in the text.  }

\begin{tabular}{lcccc}\hline
Reaction           & $np \rightarrow dX$ & $np \rightarrow d\pi\pi$
& \multicolumn{2}{c}{$np \rightarrow np$}  \\
Ref. & H-TA73\cite{H-TA73} & WASA07\cite{WASA07}
& \multicolumn{2}{c}{SP07\cite{SP07}}  \\
PW     &          &         &      $^3S_1$ & $^3D_3$ \\
 \hline\hline
Peak   &   2460\footnotemark[1] &   2410   & none?  & ?  \\
$\sigma_{{\rm tot,}J}$   &      &     &  3.6  & 3.2   \\
$\sigma_{{\rm inel,}J}$  &      &     &  0.41 & 2.3   \\
$\sigma_{d \pi\pi,J}$     & 0.63 & 2.3 & $\;\approx 0.12\;$
& $\;\approx 0.8\;$ \\
$\Gamma$   & $ > 220$\footnotemark[1] & $< 100$ &  160\footnotemark[2] &  160\footnotemark[2] \\
\hline

\end{tabular}
\footnotetext[1]{Theoretical calculation of \cite{BarNir75}.}
\footnotetext[2]{From Eq.(\ref{GammaInel}) at 2410 MeV.}
\end{table}

The energy dependence of the production cross section $\sigma_{d\pi\pi}$
(all final pion states) has been measured. Two rough fits to different
data are shown in Table~\ref{tab:ABC}. The recent preliminary WASA07
results \cite{WASA07} agree roughly with the older
Heidelberg-Tel-Aviv data (H-TA73) \cite{H-TA73}. Information can also
be deduced from $NN$ scattering. For comparison, the table gives the 
PW total and inelastic (or reaction) cross sections from the latest
energy-dependent solution SP07 \cite{SP07}. We must next estimate the
fraction of $\sigma_{\rm inel}$ that goes through the $d\pi\pi$ channel.

The estimate is made by using the following two assumptions: (1) The
single-pion production cross section in either $J$ state is
certainly not zero because a pion can be produced with the dibaryon
left in isovector states. However, in order to maximize our estimate
for $\sigma_{d\pi\pi}$, we shall ignore the contributions of all
one-pion decay branches. (2) The experimental cross section
$\sigma_{d\pi^+\pi^-}$ ($\sigma_{np\pi^+\pi^-}$) is known to be
0.270 (0.55) mb at $W=2340\,$MeV, and 0.33 (4.05) mb at
$W=2510\,$MeV \cite{Dubna80}. These points straddle the 2410 MeV ABC
peak of WASA07 \cite{WASA07}. Both cross sections should increase
significantly as one approaches the ABC peak from below. Above the
ABC peak, the larger dibaryon breakup cross section probably
reflects an increase in phase space, including the number of other
contributing states. We now assume that both cross sections at and
below the the ABC peak are dominated by the same resonance in one of
the two $J$ states in the table. Using the ratio 
$0.27/0.55 \approx 0.5$ of these cross
sections at $W = 2340\,$MeV for the entire resonance, we therefore
assume that $\sigma_{d\pi\pi} \approx \sigma_{\rm inel}/3$ to get
the rough and perhaps generous estimates shown in the table. Finally
the estimated width for the two $\Delta\Delta$ states is that of
Eq.(\ref{GammaInel}) which includes the reduction in phase space for
bound $\Delta$s.

Table \ref{tab:ABC} suggests that it is relatively unlikely that the
ABC effect originates from a dibaryon resonance in the $^3S_1^{NN}$ 
channel. The main reason is that the $NN$ scattering described by SP07 
is highly elastic so that $\sigma_{d\pi\pi}$ is far too small.  The 
situation for the $^3D_3^{NN}$ channel is more promising but not 
without difficulty unless the preliminary WASA07 estimate of 
$\sigma_{d\pi\pi}$ is reduced.

Additional information can be obtained from the energy dependence of
the SP07 $^3D_3^{NN}$ scattering amplitude \cite{SP07}. They are 
non-resonant at the QDCSM3 resonance mass of 2270 MeV, but they are 
too uncertain at the ABC peak at 2410 MeV to settle the question 
of an $NN$ resonance there. 

The resonance widths are also of interest, especially for the exclusive
reaction that excludes contributions from three and more pions on 
the high-energy side of the possible resonance. The theoretical calculation of
Bar-Nir, Risser and Schuster \cite{BarNir75} is based on a one-pion
exchange excitation to two $\Delta$s followed by a pion emission from each
$\Delta$. Their calculated resonance width is close to free-space value
$2\Gamma_{f\Delta} = 228\,$MeV used in their calculation. Our estimated
decay width shown in Table~\ref{tab:ABC} is much smaller but not as small
as the preliminary WASA07 value. As for the branching ratio $B_{NN}$ given
in Table~\ref{tab:ResonanceSummary}, the calculated value for our quark
models seems too small, but the experimental value from SP07 at 2400 MeV
is not necessarily reliable.

In his study of the ABC effect, Alvarez-Ruso \cite{A-Ruso99} has pointed
out that the $\Delta\Delta$ contribution is greatly reduced when short
range repulsive correlations are included in the $NN$ channels. Then
the cross section at $W = 2240\,$MeV, some distance below the ABC peak,
is found to be dominated by the $NN^*(1440)$, with both pions emitted
by the decaying Roper resonance. However, at this lower energy region,
the SP07 $np\,^3D_3$ Argand phase motion is not resonant.

Short range correlations are already included in the quark models used 
here. They are not the short-range repulsion from the exchange of vector 
mesons (specifically the isoscalar $\omega$ meson), which would reduce if 
not eliminate the $\Delta\Delta$ resonance. Our short range correlations 
come from Pauli antisymmetrization and channel coupling effects generated 
by overlapping clusters of quarks, including baryon excitations and 
hidden-color configurations that enhance rather than reduce these 
short-distance phenomena. 

The relative importance of these explicit quark effects found in our 
studies also comes from the use of large baryon clusters of quarks in 
all the quark models used here. The situation could be different if the 
baryon ``bags'' are small \cite{LittleBag} and the meson clouds around 
them are thicker. Our calculated results for the dibaryon spectrum is 
sensitive to the model nucleon size used in the QDCSM, but not in the
ChQM. The baryon spectrum on the other hand is quite sensitive to the
model nucleon size, especially in the radial excitations. Past 
calculations in the ChQM \cite{chiralmodel1} favors the choice near 
$b=0.52\,$fm. With this choice, the theoretical $^3D_3^{NN}$ resonance 
arising from the $^7S_3^{\Delta\Delta}$ bound state appears at about 
2390 MeV in ChQM with quadratic confinement, probably at 2420 MeV with 
linear confinement, and at 2360 MeV for the QDCSM. 

Experimental $ed$ form factors at large momentum transfers that show 
the premature dominance of six-quark effects seem to favor the kind of 
quark models studied here over the more traditional short-range $NN$ 
repulsive correlation \cite{CHH97}. The experimental confirmation of a 
$NN\,^3D_3$ resonance would be a dramatic demonstration of quark 
effects in the resonance region. Its experimental non confirmation on 
the other hand would point to a missing short-range repulsion in our 
quark models, a repulsion that is usually attributed to vector meson 
exchanges in traditional meson exchange models of nuclear forces.

\section{Conclusion}

We have studied resonances in $NN$ scattering in a theoretical treatment 
of two baryon clusters of quarks interacting by Pauli antisymmetrization
and by gluon and pion exchanges. The nucleons can resonate by changing 
into $\Delta$s, but only if the resulting baryons attract each other
with sufficient strength to stay below its $S$-wave breakup threshold. 
The absence of $NN$ $S$-wave resonances in the SP07 partial-wave 
amplitudes places an approximate upper bound of $b < 0.53\,$fm on 
the nucleon size in the QDCS quark model. This restriction in turn 
requires that the $^3D_3^{NN}$$+$$^7S_3^{\Delta\Delta}$ 
resonance mass should exceed $2340\,$MeV.
In ChQMs, the $NN$ system does not resonate in relative-$S$ waves, but 
it has a $^3D_3$ resonance at 2390-2420 MeV. This $^3D_3$ resonance 
is thus a promising candidate for the explanation of the ABC structure at 
2410 MeV in the production cross section of the reaction 
$pn$$\rightarrow $$d\pi\pi$. 

The most promising isovector $NN$ resonance candidate in our quark 
models appears in the $^1D_2^{NN}$ state and comes from a bound or almost 
bound $^5S_2^{\Delta\Delta}$ state. None of the quark models used has 
bound $N\Delta$ $P$-states that might generate odd-parity isovector resonances.

It is satisfying that these simple quark models containing only a few 
adjustable parameters fitting the $N, \Delta$ masses and the deuteron 
binding energy can yield physically interesting information about the 
possibility of dibaryon resonances at the much higher energies near the 
$\Delta\Delta$ and $N\Delta$ thresholds of $NN$ scattering. Their success 
is partly due to the fact a good part of the available collision energy 
in the center of mass frame has been used to excite the nucleons into 
one or two $\Delta$s whose mass is fitted by the models. The system is 
thus effectively at much lower energies in these $\Delta$ channels. 
The success is also derived from the ability of these simple models to 
capture some essential features of the baryon-baryon interactions in 
different energy regimes in the many channels involved in the calculation. 

The quark models used have many shortcomings. In the context of 
nonrelativistic models alone, a quantitative fit to $NN$ phase shifts 
appears difficult without the fine tuning provided by the addition of 
the many meson exchange terms that appear in conventional boson-exchange 
potentials~\cite{Machleidt89,MS01}. The need to use both quarks and 
meson exchanges suggests that the resonance region of interest 
in this paper is a transition region between the traditional low-energy 
regime of baryons interacting via meson exchanges and the high-energy 
regime of quarks interacting by quantum chromodynamics.  

From a more technical perspective, our theoretical description can be 
improved by treating the $\Delta$s as decaying particles. 
It would be difficult to go beyond this improvement because explicit 
pion channels contain three or four bodies. In our discussion of the ABC 
effect, it is of considerable interest to improve upon our very rough 
estimate of the partial-wave $pn$$\rightarrow $$d\pi\pi$
production cross sections from $NN$ partial-wave amplitudes. In spite of 
these limitations, it is clear that additional experimental knowledge and 
theoretical studies of $NN$ properties in the $NN$ resonance region near 
the $N\Delta$ and $\Delta\Delta$ thresholds will add significantly to our 
understanding of quark dynamics between baryons.   

Finally we should add that the final report of the CELSIUS-WASA Collaboration 
on the ABC effect has now appeared~\cite{WASA08}. The structure in the
total cross section for the $pn \rightarrow d\pi^+\pi^-$ reaction centers at 
2.39 GeV with a width of 90 MeV. 

This work is supported by the NSFC grant 10375030, 90503011,
10435080.

\end{document}